\pdfoutput=1
\documentclass[prx,twocolumn,showpacs,amsmath,amssymb,floatfix]{revtex4-2}

\usepackage{graphicx}
\usepackage{dcolumn}
\usepackage{bm}
\usepackage[colorlinks=true, linkcolor=blue, urlcolor=blue,
citecolor=blue]{hyperref}

\begin{document}

\title{Comment on ``Spin-orbit coupling induced ultrahigh-harmonic generation from magnetic dynamics'' with prescriptions on how to validate scientific software  for computational quantum transport}
\author{Branislav K. Nikoli\'{c}}
\email{bnikolic@udel.edu}
\affiliation{Department of Physics and Astronomy, University of Delaware, Newark, DE 19716, USA}
\author{Jalil Varela-Manjarres}
\affiliation{Department of Physics and Astronomy, University of Delaware, Newark, DE 19716, USA}

\begin{abstract}
In a recent paper [Phys. Rev. B {\bf 105}, L180415 (2022)], Ly and Manchon used  open source code    
{\tt TKWANT} for time-dependent computational quantum transport to predict surprising features in the mature field of current pumping by magnetization dynamics in spintronics---in the presence of spin-orbit (SO) coupling, the pumped charge current oscillates at both the frequency $\omega_0$ of magnetization precession and high harmonics $N=\omega/\omega_0$,  reaching {\em astonishingly high} cutoff \mbox{$N_\mathrm{max} \simeq 1000$} by increasing the SO coupling. This prediction could also open new avenues for applications as such currents would emit electromagnetic radiation ``deep in the terahertz regime'' ({\em op. cit.}). However,  results in the paper  violate two basic theorems of time-dependent quantum transport: ({\em i}) current response to time-periodic external field {\em must be perfectly periodic} itself in the long time limit for a two-terminal device because its active region is attached to two semi-infinite leads bringing continuous energy spectrum; and ({\em ii}) no DC component of charge current is allowed in the left-right symmetric two-terminal devices, or in asymmetric devices its value cannot be changed by simply increasing the SO coupling. We illustrate these two theorems by using completely different calculations applied to one-dimensional two-terminal devices with either ferromagnetic (for which the device is left-right symmetric) and antiferromagnetic (for  the  device is left-right asymmetric) active region hosting the Rashba SO coupling. We conclude that harmonics in pumped current in the presence of SO coupling do exist, but their ``ultrahigh'' cutoff is an artifact of either ``bugs'' or inadequate algorithms selected within {\tt TKWANT}. Finally,  we suggest strategies for {\em validating} time-dependent  quantum transport  codes, or selection of  algorithms by a user within putatively validated (by developers) code,  prior to deploying them to produce research papers. 
\end{abstract}

\maketitle

\begin{figure*}
	\centering
	\includegraphics[scale=0.94]{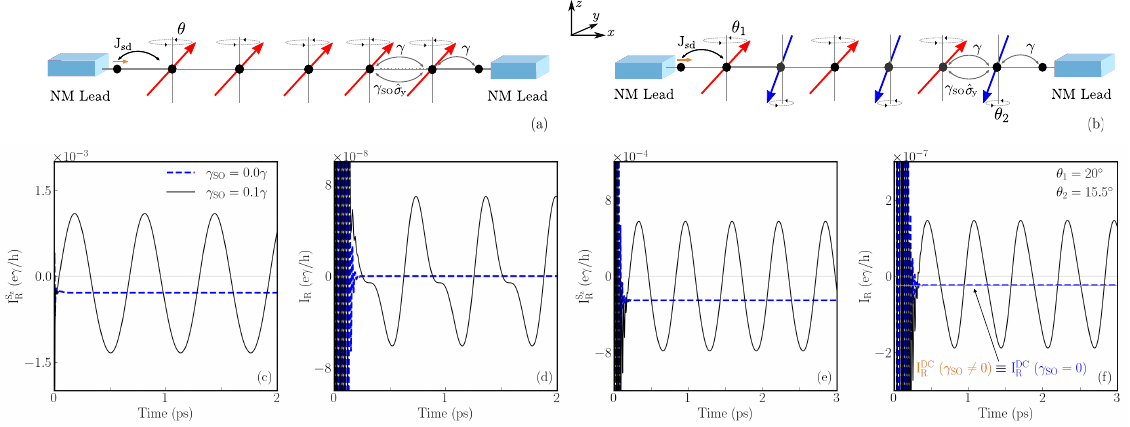}
	\caption{(a) Schematic view of a ferromagnetic metal, modeled on 1D TB lattice of $N=9$ sites (five ferromagnetic sites are shown in the illustration), which is  attached to two NM leads modeled as semi-infinite TB chains. Its classical localized magnetic moments \mbox{$\mathbf{M}_i(t)=\big(\sin \theta \cos(\omega_0 t),\sin \theta \sin(\omega_0 t), \cos \theta \big)$}  (red arrows) are steadily precessing with frequency $\hbar \omega_0=0.01 \gamma$ and  precession cone angle  $\theta=20^\circ$ around the $z$-axis due to assumed resonant absorption of microwaves. The conduction electrons within the chain interact with $\mathbf{M}_i$ via $sd$ exchange interaction of strength $J_{sd}=0.1 \gamma$ and they also experience the Rashba SO coupling~\cite{Manchon2015}, so that their Hamiltonian is given by $\hat{H}(t)  =  - \gamma \sum_{\langle ij \rangle}  \hat{c}_{i}^{\dagger}  \hat{c}_{j} - i\gamma_\mathrm{SO} \sum_{\langle ij \rangle}  \hat{c}_{i}^{\dagger}  \hat{\sigma}_y \hat{c}_{j} - J_\mathrm{sd}\sum_i\hat{c}_i^\dagger \hat{\bm \sigma} \cdot \mathbf{M}_i(t) \hat{c}_i$. Here    \mbox{$\hat{c}_i^\dagger=(\hat{c}_{i\uparrow}^\dagger \  \ \hat{c}_{i\downarrow}^\dagger)$} is a row vector containing operators $\hat{c}_{i\sigma}^\dagger$ which create an electron with spin $\sigma=\uparrow,\downarrow$ at site $i$; $\hat{c}_i$ is a column vector containing the corresponding annihilation operators;  $\gamma$ is the hopping parameter between  nearest-neighbor  sites; $\gamma_\mathrm{SO}$ is an additional spin-dependent hopping~\cite{Nikolic2006} due to the Rashba SO coupling~\cite{Manchon2015}; and  \mbox{$\hat{\bm  \sigma} = (\hat{\sigma}_x,\hat{\sigma}_y,\hat{\sigma}_z)$} is by the vector of the Pauli matrices. In the absence of any bias voltage between macroscopic reservoirs into which NM leads terminate, dynamics of localized magnetic moments $\mathbf{M}_i(t)$  drives conduction electrons  out of equilibrium so that they comprise pumped  (b) spin $I^{S_z}_p(t)$ [other two components $I^{S_x}_p(t)$ and $I^{S_y}_p(t)$ are also nonzero but not shown] and (c) charge $I_p(t)$ currents plotted in the right NM lead $p=R$. At $t=0$, electrons are described by the grand canonical equilibrium density matrix~\cite{Petrovic2018} and $\mathbf{M}_i$ start to precess leading to transient currents at early times in panels (b) and (c), while perfectly periodic [or constant~\cite{Tserkovnyak2005}, like blue dashed lines at zero SO coupling] currents are established at sufficiently long times. Panels (d)--(f) are counterparts of panels (a)--(c) for antiferromagnetic metal of $N=10$ sites [six antiferromagnetic sites are shown as the illustration in (d)].}
	\label{fig:fig1}
\end{figure*}

In Ref.~\cite{Ly2022}, the authors conduct time-dependent computational  quantum transport study of charge current pumping by precessing magnetization in the presence of spin-orbit (SO) coupling. The study is a welcome addition to more than two decades of intense efforts~\cite{Tserkovnyak2005,Ando2014} in spintronics where spin or charge current pumping by dynamical localized magnetic moments~\cite{Petrovic2018} is one of the ubiquitous quantum transport phenomena occurring even at room temperature.  Although classic review paper~\cite{Tserkovnyak2005} already noticed that widely-used scattering approach~\cite{Brouwer1998} to adiabatic quantum pumping applied to this problem needs substantial upgrade to handle presence of SO coupling (see quote on p. 1397 of Ref.~\cite{Tserkovnyak2005}:``Strong spin-orbit coupling immediately at interfaces, for example, requires generalization of spin-pumping and circuit theories beyond the scope of this review.''), only a handful of studies~\cite{Mahfouzi2012,Chen2015,Ciccarelli2015,Ahmadi2017,Dolui2020a,Dolui2021} have looked at this issue while being focused on the DC component of pumped 
currents that is usually detected experimentally~\cite{Ando2014}. In contrast, Ly and Manchon~\cite{Ly2022} performed real-time evolution of a two-terminal  device , via open source package {\tt TKWANT}~\cite{Kloss2021}, where its ferromagnetic metal (FM) or antiferromagnetic metal (AFM) as the active region hosts steadily precessing classical localized magnetic moments  at frequency $\omega_0$ while its conduction electrons are subjected to the Rashba SO coupling~\cite{Manchon2015}. This setup is akin to one-dimensional (1D) illustration in Fig.~\ref{fig:fig1}(a), but using two-dimensional (2D) tight-binding (TB) square lattice in Ref.~\cite{Ly2022}. The authors find that the fast Fourier transform (FFT) of pumped charge current  $I_p(t)$ into normal metal (NM) lead $p=L,R$ ($L$-left; $R$-right) contains peak at frequency $\omega_0$, as well as peaks at higher frequencies $\omega=N\omega_0$, where $N$ is both even and odd integer. The cutoff for such high harmonics of pumped current can reach~\cite{Ly2022} ``ultrahigh'' order $N_\mathrm{max} \simeq  1000$.

The presence of high harmonics in oscillatory functional dependence of time for both spin and charge currents (note that Ref.~\cite{Ly2022} presents results only for charge current) pumped by precessing magnetization in the presence of SO coupling has been independently confirmed~\cite{Manjarres2021} by completely different computational strategies, such as: by  time-dependent quantum transport code~\cite{Petrovic2018} different than {\tt TKWANT}; or by {\em time-independent}  calculations~\cite{Manjarres2021} based on the Floquet formalism~\cite{Mahfouzi2012,Dolui2020a,Moskalets2002,Moskalets2011,Arrachea2006}. However, the claim of ``ultrahigh'' value \mbox{$N_\mathrm{max} \simeq 1000$}  seems {\em implausible}. For example, study~\cite{Lysne2020} of high harmonics of charge and spin currents pumped by light-driven 2D electron systems with the Rashba SO coupling, which is mathematically analogous problem (aside from time-dependence in Ref.~\cite{Ly2022} being in the diagonal elements of TB Hamiltonian while it is in the off-diagonal elements in Refs.~\cite{Lysne2020}) of periodically-driven noninteracting SO-split electronic system, finds only $N_\mathrm{max} \simeq 40$. In nearly identical problem of 2D electrons on a honeycomb lattice with precessing magnetic moments,  only $N_\mathrm{max} \simeq 25$ is found~\cite{Manjarres2021} with increasing Rashba SO coupling. 

A close inspection of Fig.~3 in Ref.~\cite{Ly2022} reveals that computed charge current is {\em manifestly not periodic}, with deviation from periodic function and sudden onset of high frequency embedded signal becoming more conspicuous with increasing SO coupling. This suggests numerical artifacts in the code or its algorithms selected as one increases SO coupling. While nonperiodic currents  pumped by time-dependent quantum systems are sometimes reported in the literature (see, e.g., Fig.~1(d) in Ref.~\cite{Imai2020}), this is invariable related to the usage of a finite-size quantum system with discrete energy spectrum. Once semi-infinite NM leads are attached---so that the whole system active region + NM leads has continuous energy spectrum, which plays a key role in  in nonequilibrium quantum statistical mechanics as it provides effective dissipation~\cite{Giuliani2012,Brataas2008}---perfectly periodic spin  \mbox{$I^{S_\alpha}_p(t)=I^{S_\alpha}_p(t+2\pi/\omega_0)$} and charge \mbox{$I_p(t)=I_p(t+2\pi/\omega_0)$} currents are ensured in the long time limit. This is illustrated by Fig.~\ref{fig:fig1}(b) [Fig.~\ref{fig:fig1}(e)] for  spin and by Fig.~\ref{fig:fig1}(c)  [Fig.~\ref{fig:fig1}(f)] for charge currents pumped into semi-infinite right (R) NM lead of 1D FM [AFM] systems in Fig.~\ref{fig:fig1}(a)  [Fig.~\ref{fig:fig1}(d)] we choose as examples. Thus, the authors citing Ref.~\cite{Imai2020}, where no leads and continuous energy spectrum were used, to justify nonperiodic features of their currents is due to misunderstanding of the properties of the result that {\tt TKWANT} {\em must} produce when properly employed on harmonically driven quantum systems attached to semi-infinite leads.

The second theorem violated by the results of Fig.~3 in Ref.~\cite{Ly2022} is more subtle and has to do with the {\em key requirement}
to obtain the DC component, \mbox{$I_p^\mathrm{DC} =\frac{1}{\tau} \int_0^\tau I_p(t) dt \neq 0$} where \mbox{$\tau = 2 \pi/ \omega_0$} is one period, of pumped charge current. Note that 
\begin{eqnarray}\label{eq:sum}
I_p(t) & = &  I_p^\mathrm{DC} + \sum_{N=1}^\infty [(I_{p,N}+I_{p,-N}) \cos(N\omega_0t) \nonumber \\
\mbox{} && + i(I_{p,N}-I_{p,-N}) \sin(N\omega_0t)],
\end{eqnarray}
where $|I_{p,N}|$ is the $N$th harmonic of charge current flowing into lead $p$ that was extracted from FFT of $I_p(t)$ in Ref.~\cite{Ly2022}. The number of non-negligible  harmonics and their magnitude can be alternatively, as well as more accurately~\cite{Manjarres2021}, obtained from the Floquet scattering matrix~\cite{Moskalets2002,Moskalets2011,Arrachea2006,Manjarres2021}. For $I_p^\mathrm{DC} \neq 0$ to happen, {\em the left-right symmetry of a two-terminal device must be broken}~\cite{Vavilov2001,Moskalets2002,FoaTorres2005,Bajpai2019}. For example, if the left-right symmetry is broken by both inversion and time-reversal symmetries being violated dynamically, such as by standard example of two spatially separated potentials oscillating out-of-phase~\cite{Brouwer1998}, one finds $I_p^\mathrm{DC} \propto \omega_0$ (termed adiabatic pumping) in the low frequency regime. In contrast, if only one of those two symmetries is broken, and this does not have to occur dynamically,  then $I_p^\mathrm{DC} \propto \omega_0^2$, termed nonadiabatic pumping~\cite{Vavilov2001,Moskalets2002,FoaTorres2005,Bajpai2019} (e.g., just by adding nonzero on-site potential at one site of the device in Fig.~\ref{fig:fig1}(a) would statically break~\cite{Bajpai2019} the left-right symmetry and produce $I_p^\mathrm{DC} \propto \omega_0^2$~\cite{Chen2009}). Thus, perfectly left-right symmetric device in Fig.~\ref{fig:fig1}(a) is expected to exhibit $I_p^\mathrm{DC} \equiv 0$, as confirmed by Fig.~\ref{fig:fig1}(c) both in the absence or presence of SO coupling.

Figure 3(a) of Ref.~\cite{Ly2022} for very small SO coupling does show $I_p^\mathrm{DC} = 0$ [i.e., drawing horizontal line at zero shows that oscillatory current in Fig. 3(a) is symmetric with respect to this line]. But the authors use AFM in which case one actually expects $I_p^\mathrm{DC} \neq 0$, as confirmed by our example in Fig.~\ref{fig:fig1}(f). This is because AFM depicted in Fig.~\ref{fig:fig1}(d) breaks the left-right symmetry [there is red LMM on the left edge and blue LMM on the right edge of the active region in Fig.~\ref{fig:fig1}(f)]. Further problem with Fig.~3 of Ref.~\cite{Ly2022} is that $I_p^\mathrm{DC} \neq 0$ turns nonzero in  Fig.~3(b)--(d) with increasing SO coupling while changing its magnitude for different values of SO coupling in panels (b) to (d). Since  $I_p^\mathrm{DC} \neq 0$ means finite transmitted charge into the external circuit, which can do useful work as an example of photovoltaic effect (i.e., generation of DC current by radiation of a finite frequency~\cite{Vavilov2001}), Fig.~3(b)--(d) are essentially reporting violation of the conservation of energy, suggesting the trouble with the code. The correct calculation on any AFM system must produce  $I_p^\mathrm{DC} \neq 0$ that is insensitive to the value of the SO coupling, as confirmed by the example in Fig.~\ref{fig:fig1}(f) where we denote explicitly $I_p^\mathrm{DC}(\gamma_\mathrm{SO} \neq 0)  \equiv I_p^\mathrm{DC}(\gamma_\mathrm{SO} = 0)$.

In general, the origin of two violations produced by time-dependent computational quantum transport employed in Ref.~\cite{Ly2022} could be either due to ``bugs'' in the software or improper selection of its algorithms. For example, time-dependent currents in {\tt TKWANT} are obtained by integrating over energy an expression involving auxiliary wavefunctions~\cite{Gaury2014},  where  bound states have to located~\cite{Istas2018} on the energy axis and their contributions added manually~\cite{Kloss2021,tkwantmanual}. Ref.~\cite{Ly2022} does mention that ``evaluation of these integrals constitutes the most demanding task of the calculations'' but the issue of bound state contributions is never discussed. These states do not contribute to current in the absence of SO coupling, buy they start to contribute~\cite{Kloss_private} to it once the SO coupling is switched on. Since this is precisely how Fig.~3 of Ref.~\cite{Ly2022} starts to violate two theorems, i.e., with increasing SO coupling, it is possible that {\tt TKWANT} is ``bug''-free but the authors of Ref.~\cite{Ly2022} have simply not included all  contributions from bound states and/or converged energy integrals~\cite{tkwantmanual} (whose integrands can be highly spiky function~\cite{Gaury2014}). Additional pitfalls one can encounter when performing ${\tt TKWANT}$ simulations are discussed in its manual~\cite{tkwantmanual}. 

We note that the need to verify and validate complex scientific software has been recognized long ago~\cite{Hatton1997,Post2005}, and the software and validation procedures have become only more intricate since then~\cite{Lejaeghere2016,Vogel2019}. This includes a possibility that combination of different modules within the software could be wrong even if individual modules are well-tested. Otherwise blind trust in the code can lead to huge financial losses, as exemplified by seismic codes motivating drilling for oil in incorrect spots or computationally designed aircrafts that fail testing in  real-world wind tunnels. While it is highly unlikely that improperly validated time-dependent quantum transport codes will generate huge financial losses for nanoelectronic or spintronic industry, they could lead to proliferation of apparently fantastic but incorrect results in the physics literature where the arguments with the referees will 
be settled by saying ``we believe in the code'' instead of empirical testing and rigorous attention to details at the core of the success of modern science~\cite{Strevens2020}. 

When experimental data are available, they offer the most effective validation strategy~\cite{Post2005}. When both experimental data and exactly solvable examples are absent, one usually resorts to validation based on ``code benchmarking'' where the results of many different codes for a single problem are compared in order to determine the reasons for divergence~\cite{Post2005}, as pursued in computational materials  science~\cite{Lejaeghere2016,Vogel2019}.  Here we propose validation strategy tailored for time-dependent quantum transport codes:

\begin{itemize}
\item Using simple example, such as 1D TB chain with time-dependent on-site potentials, confirm that code results satisfy two basic theorems explained above, as illustrated by validation performed in Fig.~4 of Ref.~\cite{Bajpai2019}.

\item Compare code with an example with the {\em exact analytical} solution. This typically means focusing on quantum system driven by time-periodic external fields. For the case of spin pumping, such an example is provided by a single precessing spin within an infinite 1D TB chain whose analytical formula~\cite{Chen2009} for pumped spin and charge currents has been reproduced (see, e.g., Fig.~6 in Ref.~\cite{Petrovic2018}) in the process of validation of the code employed in Fig.~\ref{fig:fig1}, as well as by using {\tt TKWANT}~\cite{Kloss_private}. 
 
\item However, exact analytical solution of Ref.~\cite{Chen2009} cannot be generalized in the presence of SO coupling. Since SO coupling brings new physics and thereby demand on the selection of proper algorithms, code used in Fig.~\ref{fig:fig1} or code used in Fig.~3 of Ref.~\cite{Ly2022}  could be validated  for nonzero SO coupling $\gamma_\mathrm{SO}$ by time-independent numerical calculations within the Floquet formalism~\cite{Moskalets2011} where {\em numerically exact}  solution is easily achieved by increasing the size~\cite{Manjarres2021} of truncated matrices that are otherwise infinite~\cite{Eckardt2015} in the Floquet formalism. For this purpose one can also employ {\tt KWANT} package~\cite{Groth2014} for time-independent computational quantum transport, where Floquet scattering matrix~\cite{Moskalets2002,Moskalets2011,Arrachea2006,Manjarres2021} can be extracted directly from {\tt KWANT}. We also note that finding precise  cutoff harmonic order $N_\mathrm{max}$ via time-dependent quantum transport and FFT of its results, as employed in Ref.~\cite{Ly2022}, is very difficult to achieve~\cite{Manjarres2021} due to  numerical artifacts easily introduced by the choice of time step and FFT window. On the other hand, {\tt KWANT} calculations of Floquet scattering matrix and pumped currents expressed by it can yield precise~\cite{Manjarres2021} number of high harmonics from magnetic systems with precessing magnetization.
\end{itemize}

In conclusion, we reexamined the problem of current pumping by precessing magnetization in the presence of SO coupling that was very recently explored in Ref.~\cite{Ly2022}, using open source package {\tt TKWANT}~\cite{Kloss2021} for time-dependent computational quantum transport, to predict implausibly 
large number of high harmonics in pumped current. We explain that the results of Ref.~\cite{Ly2022} violate two basic theorems of time-dependent quantum transport, and we also use simple examples [Figs.~\ref{fig:fig1}(a) and ~\ref{fig:fig1}(d)] and a completely different computational scheme~\cite{Petrovic2018} to demonstrate how the correct results [panels (b),(c),(e) and (f) of Fig.~\ref{fig:fig1}] {\em must} obey both theorems. Thus, the claim of ``ultrahigh'' harmonic number in  Ref.~\cite{Ly2022} appears to be an artifact of either ``bugs'' or incorrectly selected algorithms within {\tt TKWANT} package. Regarding the latter possibility, let us recall that in teaching of elementary Computational Physics~\cite{Giordano2006} students are exposed early to the Euler algorithm for discretization of ordinary differential equations, as well as to its failure for oscillatory problems where it generates unphysical monotonically increasing energy with time. One then learns a simple fix to the Euler algorithm by switching to the Euler-Cromer one~\cite{Giordano2006}. The same approach---looking for violation of energy conservation---is also often used to select proper algorithms for much more complicated quantum time evolution~\cite{Hatano2005}. Usage of the first of two theorems discussed above is of the same type, but the second one is far less obvious. Even if both theorems are satisfied, one should still perform additional testing of code accuracy, as exemplified by recent extensive efforts in computational materials science to compare large number of codes~\cite{Lejaeghere2016,Vogel2019}. 

For time-dependent quantum transport codes, this can be accomplished by comparing code results with an analytical formula, or with numerically exact results obtained independently of a chosen code, for one or more carefully chosen testbed systems. As suggested above, a single testbed system would not be sufficient to validate accuracy of calculations of Ref.~\cite{Ly2022} due to completely different demand on the selection of algorithms within {\tt TKWANT} for zero vs. nonzero SO coupling. A user of open source scientific software who spends time to run carefully chosen testbed example (such as the three items suggested above), instead of running software as a black box, could evade  blind reporting of invalid results due to subtle  issues~\cite{Post2005} like software components  combined could be wrong even though the components test well individually; or a software combination that is insensitive to minor component errors but still gives an invalid results.

\begin{acknowledgments}
	This work was supported by the US National Science Foundation (NSF) Grant No. ECCS 1922689.
\end{acknowledgments}


\end{document}